# STOCHASTIC MODEL OF MICROPARTICLE SCATTERING ON A CRYSTAL

# MODELO ESTOCÁSTICO DE MICROPARTICULAS DISPERSAS SOBRE UN CRISTAL

**Mikhail Batanov-Gaukhman[1]**

(1) Federal State Budgetary Educational Institution of Higher Education "Moscow Aviation Institute (National Research University)", Volokolamsk highway 4, Moscow - Russian Federation
(e-mail: alsignat@yandex.ru)

**ABSTRACT**

This paper shows that the stochastic approach to solving the problem of the scattering of elementary particles on crystal can serve as an alternative to Louis de Broglie's hypothesis on the wave properties of material particles. The results of calculations presented in this work are in good agreement with the experimentally obtained electron diffraction patterns and *X*-ray diffraction patterns and the Davisson-Germer experiments.

**RESUMEN**

Este artículo muestra que el enfoque estocástico para resolver el problema de la dispersión de partículas elementales en el cristal puede servir como una alternativa a la hipótesis de Louis de Broglie sobre las propiedades ondulatorias de las partículas materiales. Los resultados de los cálculos presentados en este trabajo concuerdan bien con los patrones de difracción de electrones y los patrones de difracción de rayos X obtenidos experimentalmente y los experimentos de Davisson-Germer.

## INTRODUCTION

In 1924, Louis de Broglie suggested that a uniformly and rectilinearly moving particle with mass m and velocity v can be associated with a plane wave

$$\psi = exp\{i(Et - \mathbf{pr})/h\}, \tag{1}$$

where $E$ is the kinetic energy of the particle; $\mathbf{p} = m\mathbf{v}$ is its momentum; $h$ is Planck's constant.

The length of such a monochromatic wave is determined by the de Broglie formula

$$\lambda_b = h/m\mathbf{v}. \tag{2}$$

This idea served as the basis for the development of wave-particle duality and, in particular, made it possible to explain a number of experiments on the diffraction of electrons, neutrons, and atoms by crystals and thin films (Davisson & Germer, 1928; Bragg, 1914). Since then, it has been assumed that the diffraction maxima in the Dewisson and Germer experiment appear in directions that meet the Wolfe-Bragg condition $2d\sin\theta_s = n\lambda_{eb}$, or taking into account the refraction of de Broglie's "electronic waves" in a crystal (Davisson & Germer, 1928):

$$2d\left(n_e^2 - \cos^2\theta_s\right)^{\frac{1}{2}} = n\lambda_{eb}, \tag{3}$$

where $d$ is the interplanar distance of the crystal lattice, $\theta_s$ is the Bragg's glancing angle (Fig. 1), $n = 1, 2, 3 \ldots$ is the order of interference (or reflection), $\lambda_{eb}$ is the de Broglie electron wavelength, $n_e$ is the refractive index of the de Broglie electron wave.

However, over the past 95 years, de Broglie waves have not been detected experimentally. They have remained an auxiliary mental construction, which allows one to describe the phenomenon mathematically, without revealing the essence of the events occurring in this case.

In the author's article based on the laws of reflection of geometric optics and probability theory, the stochastic formula (5) was derived in (Batanov-Gaukhman, 2020a) {see the formula (3.9) in arXiv:2007.13527} for calculating the volumetric diagrams of elastic scattering of microparticles (in particular, electrons) on a multilayer surface crystal.

This article shows that for certain parameters, the formula (5) {or (3.9) from [Batanov-Gaukhman, 2020a]} allows us to describe the diffraction of microparticles on crystals without using de Broglie's idea of the wave properties of matter.

The results of calculations using the formula (5) are in good agreement with the experimentally obtained electron diffraction patterns (electronograms) (see Figures 1, 6 and 7) and the results of the experiment of Davisson and Germer (1927) on the diffraction of electrons on a nickel crystal (see Fig. 8).

By "microparticles", in article [Batanov-Gaukhman, 2020a] and in this work, we mean any particles (fermions and bosons) whose sizes (or wavelength) are much smaller than the characteristic sizes of the irregularities of the reflecting surface (Kirchhoff approximation), and whose elastic reflection occurs according to the laws of geometric optics.

For example, an electron can be called a "microparticle" with an effective size of about $10^{-13}$cm, which is reflected from the surface of a crystal with characteristic sizes of irregularities greater than $10^{-11}$cm.

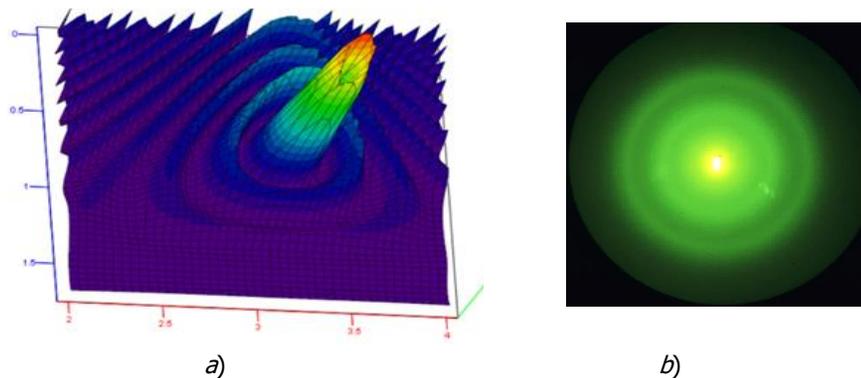

*a)*           *b)*

Fig. 1: *a*) The volumetric diagram of elastic scattering of microparticles on a multilayer crystal surface, obtained as a result of calculations by the formula (5) {or (3.9) from [Batanov-Gaukhman, 2020a]}; *b*) Experimentally obtained electron diffraction pattern (electronogram) (photo from https://www.sciencephoto.com/media/3883/view)

Also, a "microparticle" can be considered a soccer ball with a diameter of 22.3 cm, reflected from an uneven solid surface, the average radius of curvature of which is more than 20 m. "Microparticles" also include photons and phonons with a wavelength $\lambda$ by two orders less than the radius of autocorrelation of the heights of the irregularities of the reflecting surface.

Under elastic scattering of a microparticle on the surface of one of the uneven layers of the crystal, in Batanov-Gaukhman (2020a) and in this paper, means its reflection according to the laws of geometric optics (specular reflection): 1) The incident particle (or light ray), reflected particle (or light ray) and the perpendicular (normal) to the two media border in the point of particle (or light ray) incidence lie in one plane (see Figures 2 and 3); 2) The angle of incidence $Q_1$ is equal to angle of reflection $Q_2$. This phenomenon is called "specular reflection" or "elastic scattering" of microparticles.

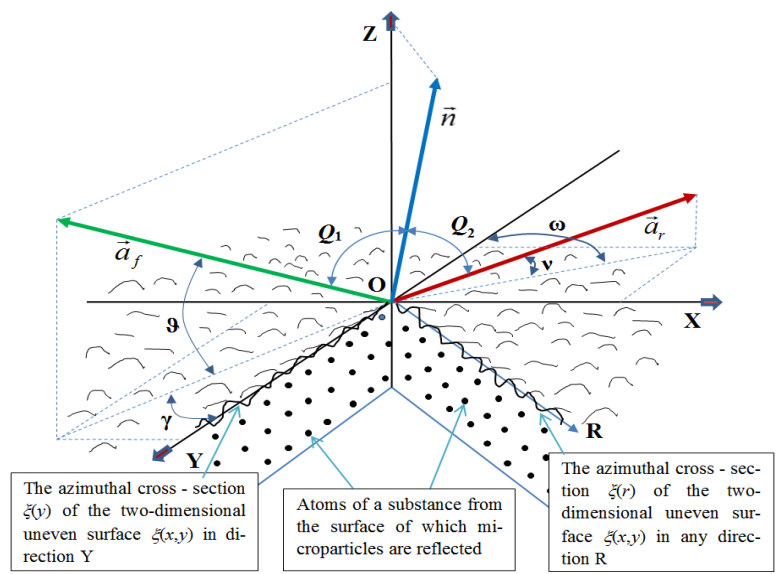

Fig. 2: Area of uneven surface (the top layer of the crystal), reflecting microparticles, where:
$\vartheta$, $\gamma$ are the angles defining the direction of microparticle incidence on a reflecting surface; $\nu$, $\omega$ are the angles defining the direction of reflection of the microparticle from this surface; **a**$_f$ is a unit vector indicating the direction to the microparticle generator; **n** is the unit normal vector to the surface at the point where the microparticle incidence; **a**$_r$ is a unit vector indicating the direction of motion of the microparticle after an elastic collision with a reflective surface

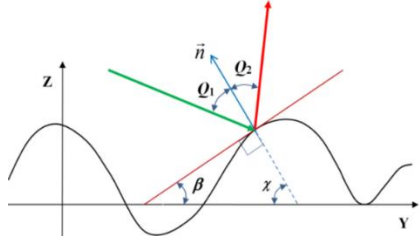

Fig. 3: Elastic (specular) reflection of a microparticle from a portion of an uneven surface according to the laws of geometric optics: 1) elastic reflection of a particle (or light ray) occurs in the plane of its incidence; 2) the angle of reflection of the particle (or ray of light) $Q_2$ is equal to the angle of its incidence $Q_1$ (i.e., the condition $Q_2 = Q_1$ is satisfied)

The object of research in this article is concentric scattering diagrams of microparticles (in particular, electrons or photons) on a multilayer crystal surface when the requirements of the Kirchhoff approximation are met.

The main aim of this work is to propose a stochastic method for calculating volumetric scattering diagrams of microparticles on a crystal without invoking to Louis de Broglie's hypothesis on the possible existence of matter waves.

**METHOD**

*Stochastic diagram of elastic scattering of microparticles on the multilayer crystal surface*

In the author's article [Batanov-Gaukhman, 2020a, arXiv:2007.13527 ], it is proposed that the irregularities of the multilayer surface of the crystal are distributed according to the multi-humped sinusoidal law

$$\rho(\xi) = \begin{cases} \dfrac{2\sin^2(\pi n_1 \xi / l_2)}{l_2} & \text{for } \xi \in [0, l_2]; \\ 0 & \text{for } \xi \notin [0, l_2], \end{cases} \quad (4)$$

where $\rho(\xi)$ is the probability density function (PDF) of the heights of the irregularities of the multilayer crystal surface (see Fig.4b); $l_1$ is the thickness of one (for example, the first) reflecting layer of the crystal (see Fig. 4); $n_1$ is the number of identical uneven layers of a sinusoidal crystal that fit in the interval [0, $l_2$]; here $l_2 = n_1 l_1$ is the depth of the multilayer crystal surface of an efficiently scattering microparticle.

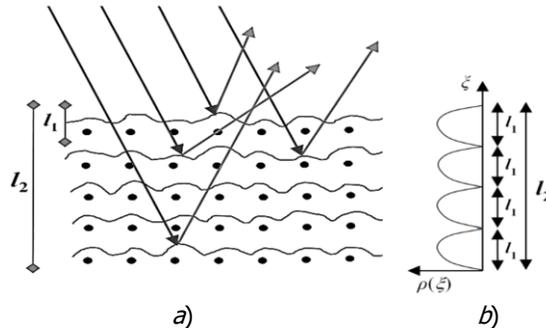

Fig. 4: *a)* Scattering of microparticles on the multilayer surface of the crystal, with each layer being considered as a separate uneven surface of the sinusoidal type; *b)* Multi-humped sinusoidal PDF (4) of the height of the irregularities of the multilayer surface of the crystal

Further, in Section 3.2 of the article [Batanov-Gaukhman, 2020a, arXiv:2007.13527], the following formula for calculating the diagrams of the elastic scattering of microparticles (DESM) on a multilayer crystal surface with the PDF (4) of the heights of irregularities is derived

$$D(v,\omega/\vartheta,\gamma) = \frac{1}{2\pi^2 l_2} \left( \frac{\cos^2(\pi n_1) - \cos(\pi n_1)\cos\left(\sqrt{\frac{a^2+b^2}{d^2}}l_2/\eta\right)}{\left(\pi n_1/l_2\right)^2 - \left(\sqrt{\frac{a^2+b^2}{d^2}}l_2/\eta\right)^2} - \frac{\cos\left(\pi n_1 + \sqrt{\frac{a^2+b^2}{d^2}}l_2/\eta\right) - 1}{\left(\pi n_1/l_2 + \sqrt{\frac{a^2+b^2}{d^2}}/\eta\right)^2} \right) \times$$

$$\times \left| \frac{d(a'_v b'_\omega - a'_\omega b'_v) + c'_v(ba'_\omega - ab'_\omega)}{d^2\sqrt{a^2+b^2}} \right|, \quad (5)$$

where, according to the notation shown in Figures 2 and 4,
$a = \cos v \cos\omega + \cos\vartheta \cos\gamma$; $\quad b = \cos v \sin\omega + \cos\vartheta \sin\gamma$; $\quad d = \sin v + \sin\vartheta$; $\quad a'_v = -\sin v \cos\omega$;
$b'_v = -\sin v \sin\omega$; $\quad c'_v = \cos v$; $\quad a'_\omega = -\cos v \sin\omega$; $\quad b'_\omega = \cos v \cos\omega$;

$$\eta = \frac{l_1^2(\pi^2 n_1^2 - 6)}{6\pi^2 r_{cor5}}, \quad (6)$$

where
$l_1$ is the thickness of one reflecting layer (i.e., the horizontal atomic plane) of the crystal (see Fig. 4);
$l_2 = l_1 n_1$ is the depth of the multilayer surface of the single crystal, effectively involved in the elastic scattering of microparticle(s) (see Fig. 4);
$n_1$ is the number of uneven layers of a single-crystal (sinusoidal type) that fit in the interval $[0, l_2]$;
$r_{cor}$ is the autocorrelation radius of one uneven layer of a sinusoidal type. This autocorrelation radius is approximately equal to the average radius of curvature of the sinusoidal irregularities of a single crystal layer;
$\vartheta, \gamma$ are the angles that specify the direction of motion of the microparticle beam incident on the crystal surface (see Fig. 2);
$v, \omega$ are the angles that specify the direction of movement of microparticles reflected from the surface of the crystal toward the detector (see Fig. 2).

As a result of calculations by formula (5) for various parameters $l_1$, $n_1$, $r_{cor}$, $\vartheta$ and $\gamma$, an infinite number of scatter diagrams are obtained, similar to two-dimensional landscapes of mountainous terrain. Some variants of these diagrams (i.e., 2D-landscapes) with different parameters $l_1$, $n_1$, $r_{cor}$ are shown in Fig. 5.

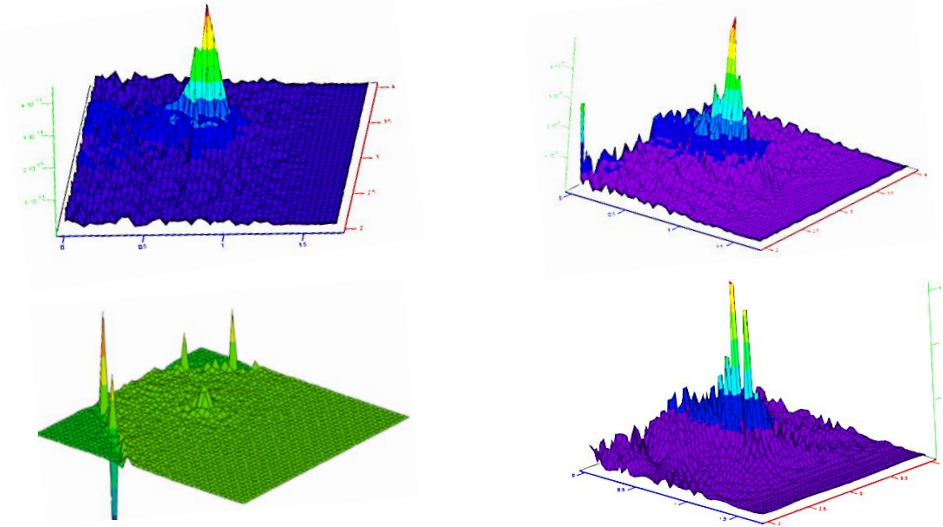

Fig. 5: The Diagrams of elastic scattering of microparticles (DESM) on a multilayer statically uneven surface of a crystal calculated by formula (5) for various parameters $l_1$, $n_1$, $r_{cor}$, $\vartheta$, $\gamma$

The entire infinite set of DESM (i.e., 2D-landscapes) obtained by calculations using formula (5), we propose to call the two-dimensional Bass-Fuchs world, in honor of Friedrich Gershonovich Bass and Joseph Moiseevich Fuks, who made a great contribution to the study scattering of waves on statistically uneven surfaces (Bass & Fuks, 1979).

**RESULTS**

*Stochastic diagrams of elastic scattering of electrons on a crystal*

Let's apply stochastic formula (5) to obtain a volumetric electron scattering diagram with characteristic dimensions of the order of ∼ 2.8×10⁻¹³ cm (Lorentz radius) on the surface of a crystal with a distance between atomic layers $h_1$ =10⁻¹¹ cm and a radius of autocorrelation of irregularities the sinusoidal surface of each atomic layer $r_{cor}$= 6×10⁻⁹ cm.

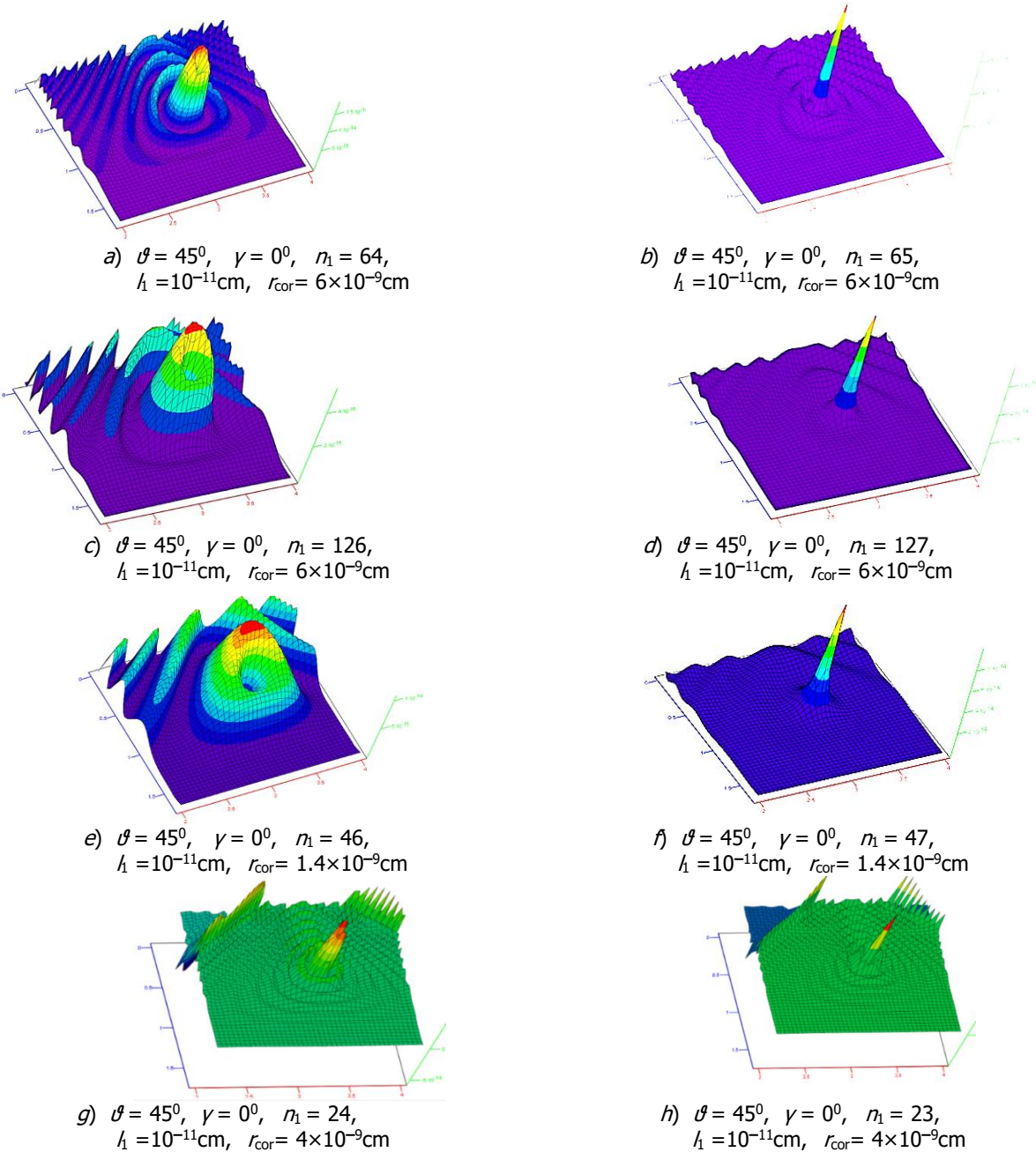

a) $\vartheta = 45^0$, $\gamma = 0^0$, $n_1 = 64$, $h_1 = 10^{-11}$ cm, $r_{cor} = 6×10^{-9}$ cm

b) $\vartheta = 45^0$, $\gamma = 0^0$, $n_1 = 65$, $h_1 = 10^{-11}$ cm, $r_{cor} = 6×10^{-9}$ cm

c) $\vartheta = 45^0$, $\gamma = 0^0$, $n_1 = 126$, $h_1 = 10^{-11}$ cm, $r_{cor} = 6×10^{-9}$ cm

d) $\vartheta = 45^0$, $\gamma = 0^0$, $n_1 = 127$, $h_1 = 10^{-11}$ cm, $r_{cor} = 6×10^{-9}$ cm

e) $\vartheta = 45^0$, $\gamma = 0^0$, $n_1 = 46$, $h_1 = 10^{-11}$ cm, $r_{cor} = 1.4×10^{-9}$ cm

f) $\vartheta = 45^0$, $\gamma = 0^0$, $n_1 = 47$, $h_1 = 10^{-11}$ cm, $r_{cor} = 1.4×10^{-9}$ cm

g) $\vartheta = 45^0$, $\gamma = 0^0$, $n_1 = 24$, $h_1 = 10^{-11}$ cm, $r_{cor} = 4×10^{-9}$ cm

h) $\vartheta = 45^0$, $\gamma = 0^0$, $n_1 = 23$, $h_1 = 10^{-11}$ cm, $r_{cor} = 4×10^{-9}$ cm

Fig. 6: The volumetric diagrams of elastic scattering of microparticles on a multilayer crystal surface, calculated by the formula (5) for various values of the parameters $\vartheta$, $h_1$, $n_1$ and $r_{cor}$

Let the electrons fall on the surface of the crystal from the direction specified by the glancing angle $\vartheta = 45^0$ and the azimuthal angle $\gamma = 0^0$, while the speed of the electrons is such that they can penetrate deep into the crystal up to the 64th layer (i.e., $n_1 = 64$).

By supplying the parameters $h_1 = 10^{-11}$cm, $r_{cor} = 6\times10^{-9}$cm, $\vartheta = 45^0$, $\gamma = 0^0$, $n_1 = 64$ in formula (5), we obtain the electron scattering diagram shown in Fig. 6a.

By slightly changing the parameters $\vartheta$, $\gamma$, $h_1$, $n_1$ and $r_{cor}$ in formula (5), we obtain the remaining electron scattering diagrams on the multilayer surface of a crystal shown in Fig. 6.

The results of calculations by formula (5) with parameters corresponding to the scattering of electrons on the multilayer surface of a crystal exceeded expectations. Among the many chaotic landscapes of the Bass-Fuchs world, regular concentric diagrams suddenly appeared, which correspond to the experimentally obtained electron diffraction patterns (see Fig. 7).

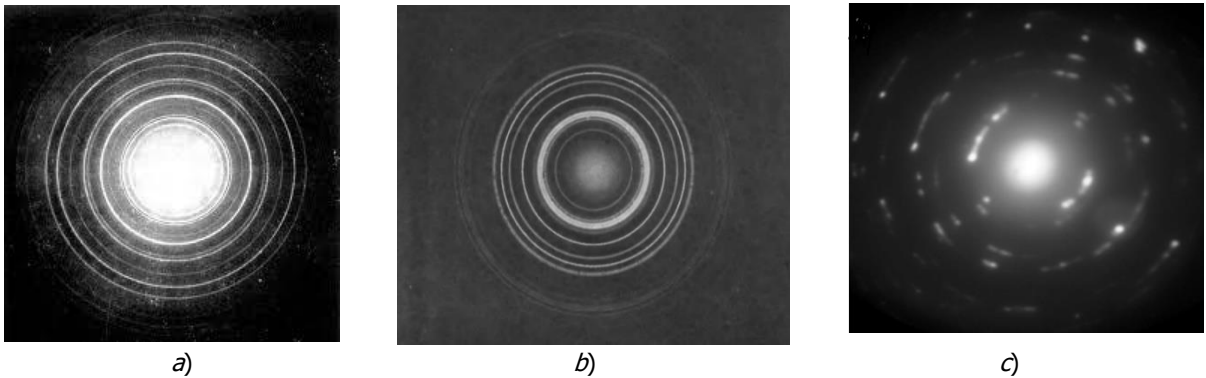

*a)*  *b)*  *c)*

Fig. 7: *a*) Electron diffraction pattern of the NaCl standard; *b*) Electron diffraction pattern of a polycrystal of hexagonal nickel hydride NiH$_2$ (http://ignorik.ru/docs/lekciya-13-eksperimentalenie-metodi-kristallofiziki.html); *c*) Electron diffraction on aluminum Al. https://www.researchgate.net/publication/295974108_Electron_Diffraction. Photos are taken from the World Wide Web in the public domain

The concentric scatter diagrams shown in Fig. 6, are obtained on the basis of the laws of geometric optics and probability theory, i.e. without involving the idea of the possible existence of de Broglie waves.

In order to make sure that the proposed stochastic interpretation of the process of scattering of microparticles (in particular, electrons) on the crystal surface is correct, let's compare the calculations by formula (5) with the results of the experiment by Davisson and Germer on the diffraction of electrons on a nickel crystal [Davisson & Germer, 1928].

*Stochastic interpretation of the Davisson-Germer experiment*

It should be expected that the number of atomic layers of the crystal surface $n_1$, onto which the incident microparticles (in particular, electrons) penetrate, mainly depends on their velocity (more precisely, the momentum $p=m\boldsymbol{v}$). In a more general case, this dependence can have the form

$$n_1 = f(p = m\boldsymbol{v}, h_1, r_{cor}, \vartheta, \gamma), \qquad (7)$$

Expression (7) can also take into account the effects of shading of a part of the deepened sections of the reflecting surface at small angles $\vartheta$, etc.

Establishment of functional dependence (7) will make it possible to more accurately match the results of calculations by formula (5) with experimental data on the diffraction of microparticles by periodic structures such as crystals and to obtain additional information on the structure of the reflecting surface.

Let's consider DESM (5) as a function of the number of layers $n_1$ of the reflecting surface of the crystal $D(n_1)$ for six fixed parameters $\vartheta$, $\gamma$, $v$, $\omega$, $h_1$, $r_{cor}$.

$$D(n_1) = \frac{1}{2\pi^2 l_2} \left( \frac{\cos^2(\pi n_1) - \cos(\pi n_1)\cos\left(\sqrt{\frac{a^2+b^2}{d^2}} l_2/\eta\right)}{\left(\pi n_1/l_2\right)^2 - \left(\sqrt{\frac{a^2+b^2}{d^2}} l_2/\eta\right)^2} - \frac{\cos\left(\pi n_1 + \sqrt{\frac{a^2+b^2}{d^2}} l_2/\eta\right) - 1}{\left(\pi n_1/l_2 + \sqrt{\frac{a^2+b^2}{d^2}}/\eta\right)^2} \right) \times$$

$$\times \left| \frac{d(a'_v b'_\omega - a'_\omega b'_v) + c'_v(ba'_\omega - ab'_\omega)}{d^2 \sqrt{a^2+b^2}} \right|,$$
(8)

where $\eta = \dfrac{l_1^2(\pi^2 n_1^2 - 6)}{6\pi^2 r_{cor}}$.

The calculation result using formula (8) as a function of $n_1$ is shown in Fig. 8a.

Taking into account the fact that the number layers of crystal penetrated by incident microparticles (in particular, electrons) depends on their velocity $n_1 = f(v)$, these calculations are in good agreement with the results of the Davisson-Germer Experiment [Davisson & Germer, 1928] (see Fig. 8b).

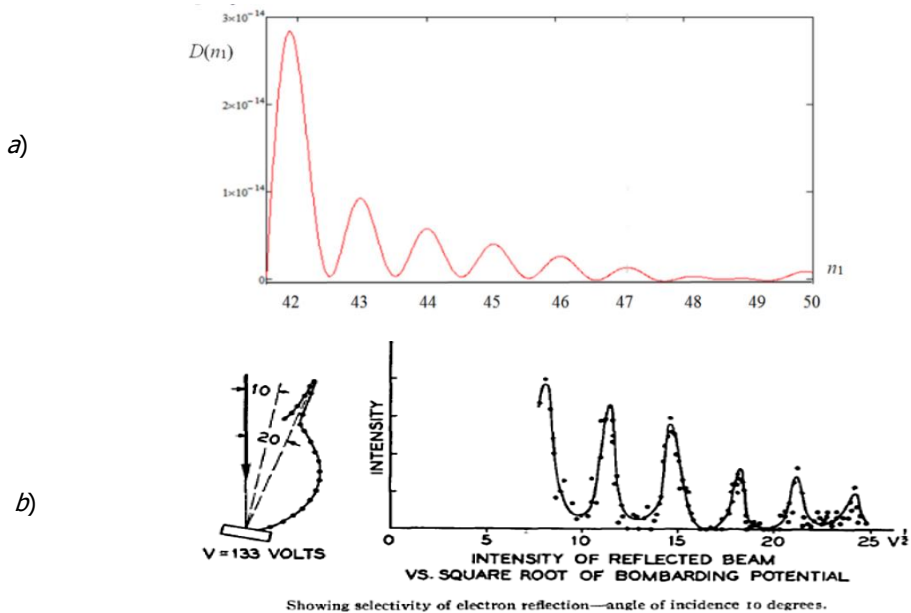

Fig. 8: *a)* The results of calculating the dependence of DURM $D(n_1)$ (8) on the number of layers $n_1$ of the reflecting surface of the crystal, which, in turn, depend on the velocity $v$ (more precisely, on the energy $E$) of microparticles incident on this surface. The calculations were performed according to formula (8) as a function of the number $n_1$, which varies in the range from 40 to 50 layers, with the following constant parameters $\vartheta = 45^0$, $\gamma = 0^0$, $v = 45^0$, $\omega = 0^0$, $h_1 = 10^{-11}$cm, $r_{cor} = 9\times 10^{-9}$cm; *b)* The intensity of an electron beam $I$ scattered on a nickel single crystal at a constant reflection angle, depending on the square root of the voltage $U$, accelerating particles in an electron gun (electron generator). This experimental dependence was first obtained in 1927 by Davisson and Germer

Formula (8) allows performing calculations in a much wider range of values of $n_1$, as shown in section 4 of § 3.2 in Batanov-Gaukhman (2020a).

Thus, formula (5), obtained in Batanov-Gaukhman (2020a, arXiv:2007.13527] on the basis of a stochastic model of scattering of microparticles on a statistically uneven multilayer crystal surface, adequately reflects the known experimental data.

At the same time, stochastic formula (5) has tangible advantages over the Wolfe-Bragg condition (3), based on the idea of the existence of de Broglie matter waves. By selecting the parameters $\vartheta$, $h$, $n_1$ and $r_{cor}$ in formula (5), one can achieve similarity with experimentally obtained electron diffraction patterns or X-ray diffraction patterns, while revealing more detailed information about the structure of the crystal.

*Even and odd number of crystal layers*

From the diagrams shown in Fig. 6, it can be seen that if an even number of layers $n_1$ is effectively involved in the reflection of microparticles, then a minimum (dip) is observed in the very center of the concentric diagram; and if the number of reflecting layers is odd, then a maximum (peak) is observed in the very center of the diagram. The same effect is observed in experiments (see Fig. 9), which once again confirms the adequacy of the proposed stochastic model of microparticle scattering on a multilayer crystal surface.

The combination of the above coincidences of the results of calculations by formula (5) with experimental data (see Fig. 6 – 9) allows us to propose a stochastic interpretation of diffraction phenomena, as an alternative to L. de Broglie's hypothesis about the existence of wave properties of matter.

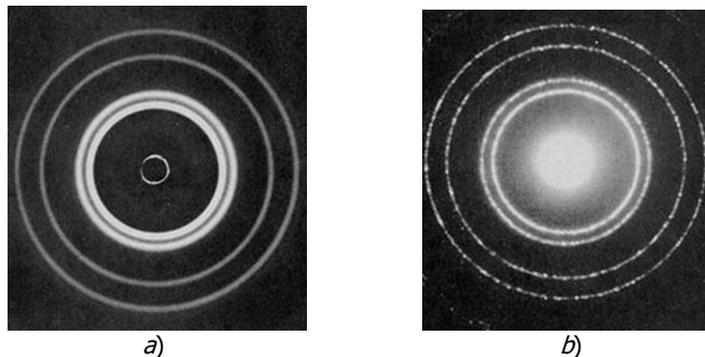

*a)*        *b)*

Fig. 9: *a)* In a number of experiments on the diffraction of microparticles, a dark spot is observed in the center of the electron diffraction pattern or X-ray diffraction pattern. *b)* In a number of other similar experiments, a bright spot is observed in the center of the electron diffraction pattern or radiograph. Photos are taken from the World Wide Web in the public domain.

However, it should be noted that for $n_1 = 4$ (i.e., with an even number of layers), not a minimum, but a maximum (peak) is observed in the center of the diagram (see Fig. 13).

*Setting the scale parameter η*

Scale parameter (6)

$$\eta = \frac{l_1^2(\pi^2 n_1^2 - 6)}{6\pi^2 r_{cor5}}$$

obtained in Batanov-Gaukhman (2020a) and Batanov-Gaukhman (2020b) when passing from the PDF of the heights of irregularities of a random process to the PDF of its derivative, taking into account the change in only the second central moment (i.e., dispersion). Higher central moments were ignored {see expressions (16) – (65) in Batanov-Gaukhman (2020b)}, since their influence in many cases is insignificant.

To compensate for this disadvantage, as well as to take into account other features of the crystal lattice, the scale parameter $\eta$ can be "adjusted" to the experimental results. For example, it can change the values of numeric constants or introduce functional dependencies on the parameters $h_1$, $n_1$ and $r_{cor}$:

$$\eta = \frac{l_1^2(\pi^2 n_1^2 - 3)}{12\pi^2 r_{cor}} \quad \text{or} \quad \eta = \frac{l_1^2(\pi^2 n_1^2 - 127)}{16\pi^2 r_{cor}} \quad \text{or} \quad \eta = \frac{l_1^2(4\pi^2 n_1^2 - 33)}{12\pi^2 r_{cor}}, \quad \text{etc.}$$

$$\eta = \frac{l_1^2[\pi^2 \cos^2(\pi n_1) - 8]}{4\pi^2 r_{cor}} \quad \text{or} \quad \eta = \frac{\ln(l_1^2)[\pi^2 tg^2(\pi n_1) - 13]}{7\pi^2 r_{cor}}, \quad \text{etc.} \qquad (9)$$

Perhaps such a "adjustment" $\eta$ will lead to a greater similarity of the results of calculations by formula (5) with real electron diffraction patterns or *X*-ray diffraction patterns. At the same time, the "tuning" of the scale parameter can make it possible to evaluate additional features of the structure and / or crystal lattice defects.

*Scattering of microparticles on one crystal layer*

In the case of scattering of microparticles on one layer of the crystal (i.e., at $n_1 = 1$), the calculation by formula (5) leads to the results shown in Fig. 10 a,b.

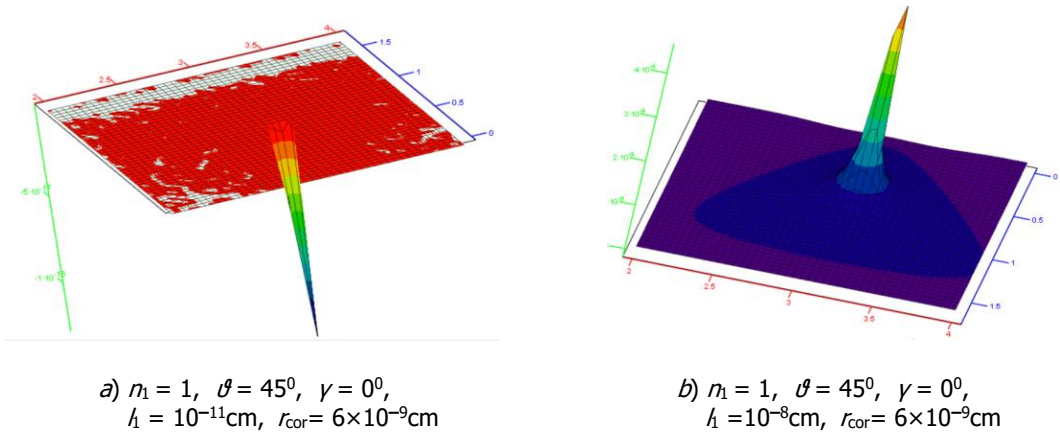

a) $n_1 = 1$, $\vartheta = 45^0$, $\gamma = 0^0$,
$h_1 = 10^{-11}$cm, $r_{cor} = 6\times10^{-9}$cm

b) $n_1 = 1$, $\vartheta = 45^0$, $\gamma = 0^0$,
$h_1 = 10^{-8}$cm, $r_{cor} = 6\times10^{-9}$cm

Fig. 10: The Diagrams of elastic scattering of microparticles (DESM) on one layer of the crystal ($n_1 = 1$), calculated by the formula (5) for various $h_1$

If the thickness of the first layer is $h_1 = 10^{-11}$cm, then the calculation result by the formula (5) is negative (see Fig. 10a). This can be explained by the fact that microparticles do not reflect from this layer but pass through it.

If this layer is thicker, for example, $h_1 = 10^{-8}$cm, then the reflection from such a layer is shown in Fig. 10b. In this case, concentric diagrams are not formed.

An interesting calculation results using the formula (5) is observed for $n_1 = 1$ and $h_1 = 4\times10^{-10}$cm (see Fig. 11). This case can be interpreted as a prediction that part of the microparticles will reflect from one layer of the crystal, and the other part of the microparticles will pass through it.

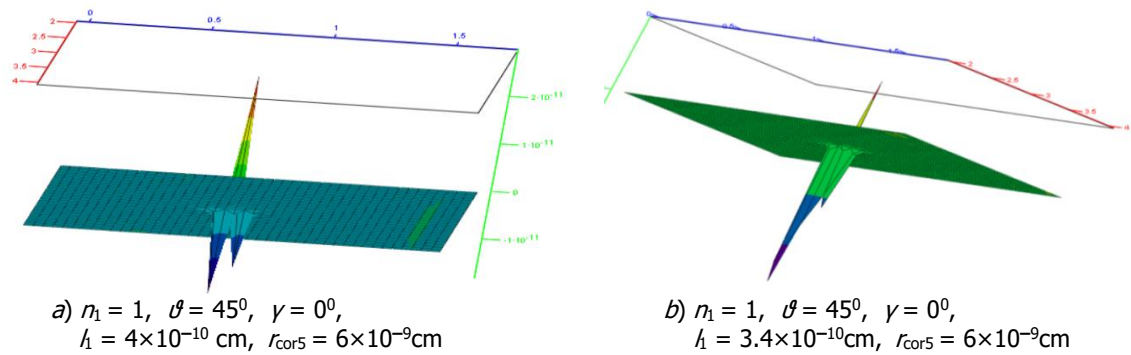

a) $n_1 = 1$, $\vartheta = 45^0$, $\gamma = 0^0$,
$h_1 = 4\times10^{-10}$ cm, $r_{cor5} = 6\times10^{-9}$cm

b) $n_1 = 1$, $\vartheta = 45^0$, $\gamma = 0^0$,
$h_1 = 3.4\times10^{-10}$cm, $r_{cor5} = 6\times10^{-9}$cm

Fig. 11: The Diagrams of elastic scattering of microparticles on one layer of the crystal,
calculated by the formula (5) for $n_1 = 1$, a) $h_1 = 4\times10^{-10}$ cm and b) $h_1 = 3.4\times10^{-10}$ cm

*Scattering of microparticles on two, three and four layers of the crystal*

The Diagrams of elastic scattering of microparticles on two, three and four layers of the crystal, calculated by the formula (5), are shown in Fig. 12 and 13.

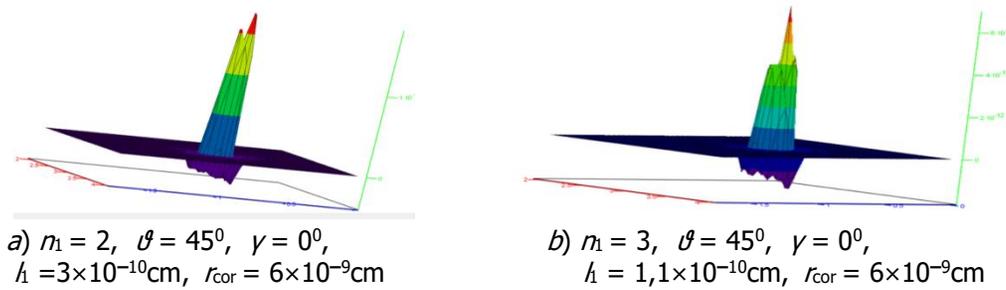

a) $n_1 = 2$, $\vartheta = 45^0$, $\gamma = 0^0$,
$h_1 = 3\times10^{-10}$cm, $r_{cor} = 6\times10^{-9}$cm

b) $n_1 = 3$, $\vartheta = 45^0$, $\gamma = 0^0$,
$h_1 = 1{,}1\times10^{-10}$cm, $r_{cor} = 6\times10^{-9}$cm

Fig. 12: The DESM on two (*a*) and three (*b*) layers of the crystal, calculated by the formula (5)

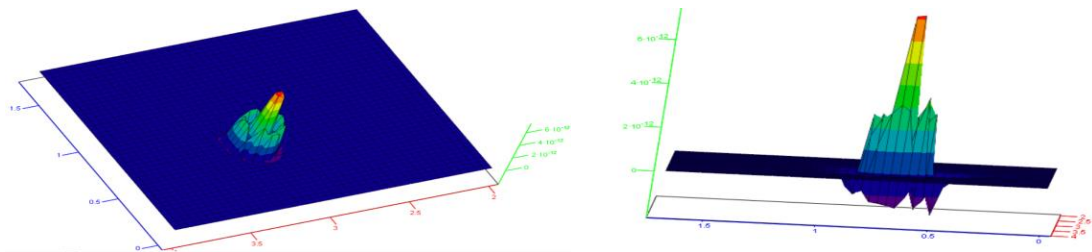

Fig. 13: The DESM on four layers of the crystal, calculated by the formula (5) for $n_1 = 4$, $\vartheta = 45^0$,
$\gamma = 0^0$, $h_1 = 1{,}2\cdot10^{-10}$cm, $r_{cor5} = 9\cdot10^{-9}$cm

*Diffraction of microparticles on the thin films*

The method for calculating the DESM, presented in [Batanov-Gaukhman, 2020a, arXiv:2007.13527], was developed on the basis that after collision with the surface of a solid, microparticles are reflected from it according to the laws of geometric optics, and do not pass through this body. But it turned out that formula (5) makes it possible to calculate the scattering diagram also for the passage of microparticles through thin films. In Fig. 14 shows the results of calculations by formula (5) of the scattering diagrams of microparticles on films consisting of 14 and 15 atomic layers.

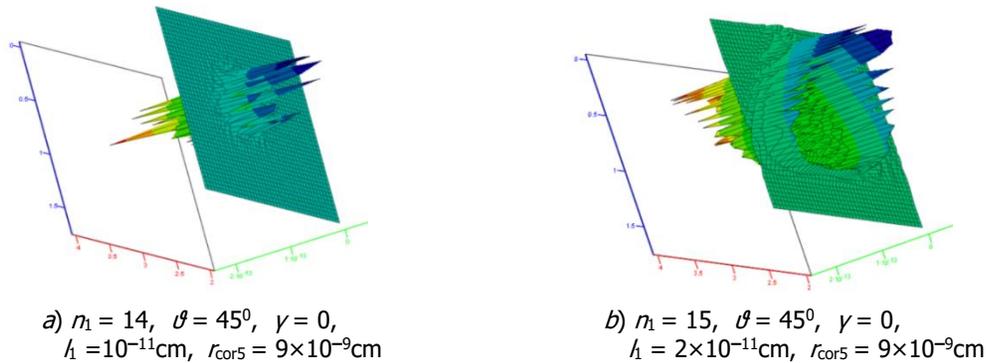

a) $n_1 = 14$, $\vartheta = 45^0$, $\gamma = 0$,
$h_1 = 10^{-11}$cm, $r_{cor5} = 9\times10^{-9}$cm

b) $n_1 = 15$, $\vartheta = 45^0$, $\gamma = 0$,
$h_1 = 2\times10^{-11}$cm, $r_{cor5} = 9\times10^{-9}$cm

Fig. 14: The diffraction maxima of microparticles passing through thin films calculated by the formula (5)

It should be noted that these diffraction maxima are obtained when microparticles fall onto thin films at angles $\vartheta$ from $25^0$ to $65^0$. In this case, some of the microparticles are reflected from uneven atoms layers of the thin film, while the other part passes through them.

When microparticles fall vertically onto the surface of the body (i.e., at $\vartheta = 90^0$), calculations using formula (5) lead to absurd results. That is, the method for calculating DESM proposed in article [Batanov-Gaukhman, 2020a] is not applicable to this case.

The fifth parameter γ

As shown above, by choosing four parameters $\vartheta$, $h_1$, $n_1$, $r_{cor}$, it is possible to achieve that the calculations by formula (5) correspond to different variants of diffraction of microparticles on a multilayer statistically uneven crystal surface.

The fifth parameter, the angle $\gamma$ (see Fig. 2), remained equal to zero in all previously considered cases ($\gamma = 0^0$).

When deriving formula (5) in Batanov-Gaukhman (2020a), it was taken into account that all azimuthal cross-sections in different directions of the homogeneous and isotropic uneven surface of the crystal are the same. Therefore, it was expected that with a change in the azimuthal angle $\gamma$, the scattering diagram should remain unchanged, and only its azimuthal direction should change.

From the diagrams shown in Fig. 15, it can be seen that at small angles $\gamma$ equal to $35^0$ and $55^0$, only the azimuthal direction of the entire diagram as a whole shifts.

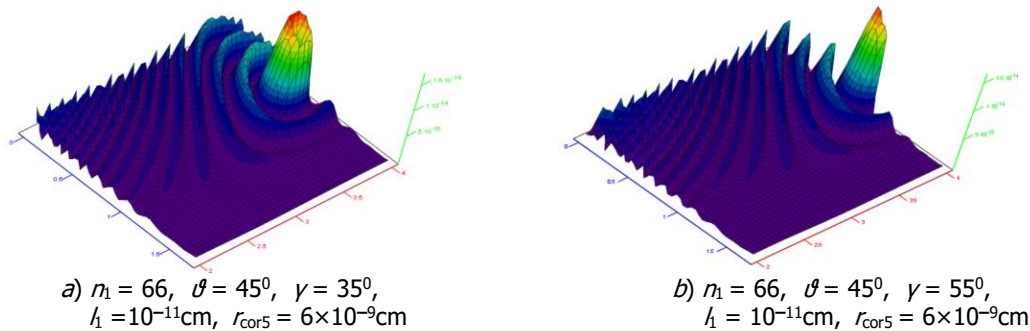

a) $n_1 = 66$, $\vartheta = 45^0$, $\gamma = 35^0$,
$h_1 = 10^{-11}$cm, $r_{cor5} = 6\times10^{-9}$cm

b) $n_1 = 66$, $\vartheta = 45^0$, $\gamma = 55^0$,
$h_1 = 10^{-11}$cm, $r_{cor5} = 6\times10^{-9}$cm

Fig. 15: The DESM on a crystal, calculated by the formula (5), for the identical $\vartheta$, $n_1$, $h_1$, $r_{cor5}$ and different angles $\gamma$

But with a further increase in the angle $\gamma$, the scattering diagram changes significantly with the remaining four parameters $\vartheta$, $h_1$, $n_1$, $r_{cor}$ unchanged (see Fig. 16).

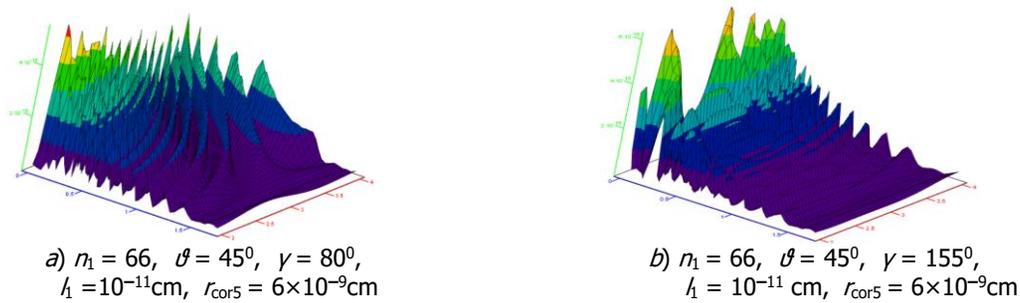

a) $n_1 = 66$, $\vartheta = 45^0$, $\gamma = 80^0$, $h_1 = 10^{-11}$cm, $r_{cor5} = 6\times10^{-9}$cm

b) $n_1 = 66$, $\vartheta = 45^0$, $\gamma = 155^0$, $h_1 = 10^{-11}$ cm, $r_{cor5} = 6\times10^{-9}$cm

Fig. 16: The DESM on a crystal, calculated by the formula (5), for the identical $\vartheta$, $n_1$, $h_1$, $r_{cor5}$ and different angles $\gamma$

At this stage of the study, it is difficult to establish whether this change is a drawback of formula (5), or it is a reflection of reality, which can be confirmed experimentally.

It can be assumed that DESM depends on the angle $\alpha$ between the projection of the azimuthal direction of motion of the incident microparticles on the *XOY* plane and the direction of the rows of atoms in the crystal lattice of the crystal (see Fig. 17).

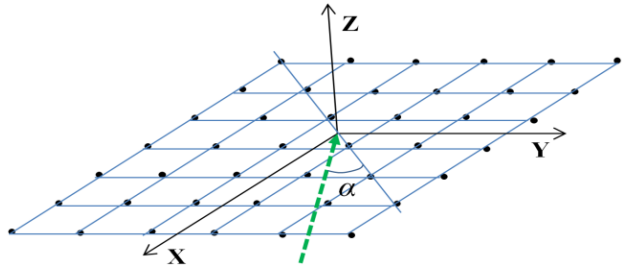

Fig. 17: The angle $\alpha$ between the projection of the azimuthal direction of motion of incident microparticles on the XOY-plane and the direction of the rows of atoms in the crystal lattice

From Fig. 17 can be seen that rotation of the plane of incidence of the microparticles at the angle $\alpha$ is accompanied by the effect of increasing the distance between the atoms of the crystal lattice, which are effectively involved in their scattering. This effect can be taken into account by increasing the correlation radius of the heights of the surface irregularities $r_{cor}$.

The scattering diagrams at $\gamma = 75^0$ and enlarged in comparison with the previous case of $r_{cor}$ and $h_1$ are shown in Fig. 18.

These calculation results by the formula (5) are subject to experimental verification. If the distortions of the DESM due to a change in the angle $\gamma$ are not experimentally confirmed, then this disadvantage can be compensated for by a change in the orientation of the reference frame.

In many cases, the coordinate axis from which the angle $\gamma$ is measured can be initially combined with the azimuthal direction of motion of the microparticles incident on the crystal surface. That is, in a number of experiments, taking advantage of the arbitrariness in choosing a reference frame, it is possible from the very beginning to achieve that $\gamma = 0^0$.

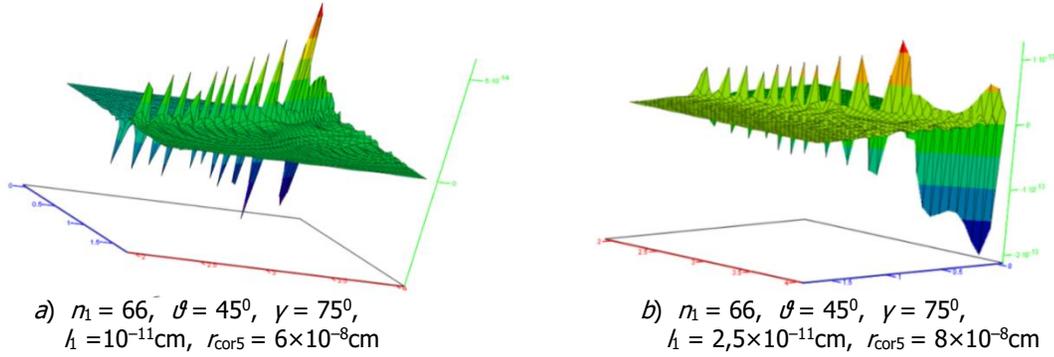

a) $n_1 = 66$, $\vartheta = 45^0$, $\gamma = 75^0$, $l_1 = 10^{-11}$cm, $r_{cor5} = 6\times10^{-8}$cm

b) $n_1 = 66$, $\vartheta = 45^0$, $\gamma = 75^0$, $l_1 = 2,5\times10^{-11}$cm, $r_{cor5} = 8\times10^{-8}$cm

Fig. 18: The DESM calculated by the formula (5), at $\gamma = 75^0$ and increased $r_{cor5}$ and $l_1$

*Scattering of microparticles on the surface of a crystal with anisotropic layers*

If each layer of the crystal has the same anisotropy, for example, of the type (2.23) in [Batanov-Gaukhman (2020a)], then in this case the following formula was obtained in [Batanov-Gaukhman,2020a, arXiv:2007.13527] for calculating the DESM on the anisotropic surface of the crystal

$$D(\nu,\omega/\vartheta,\gamma) = \frac{2}{\pi^2 l_2}\left(\frac{b^2}{a^2+b^2}\right)\left(\frac{\cos^2(\pi n_1) - \cos(\pi n_1)\cos\left(\sqrt{\frac{a^2+b^2}{d^2}}l_2/\eta\right)}{\left(\pi n_1/l_2\right)^2 - \left(\sqrt{\frac{a^2+b^2}{d^2}}l_2/\eta\right)^2} - \frac{\cos\left(\pi n_1 + \sqrt{\frac{a^2+b^2}{d^2}}l_2/\eta\right) - 1}{\left(\pi n_1/l_2 + \sqrt{\frac{a^2+b^2}{d^2}}/\eta\right)^2}\right) \times$$
$$\times \left|\frac{d(a'_\nu b'_\omega - a'_\omega b'_\nu) + c'_\nu(ba'_\omega - ab'_\omega)}{d^2\sqrt{a^2+b^2}}\right|. \quad (10)$$

where the scale parameter $\eta$ is given by Expression (6).

The results of calculations using formula (10) are shown in Fig. 19.

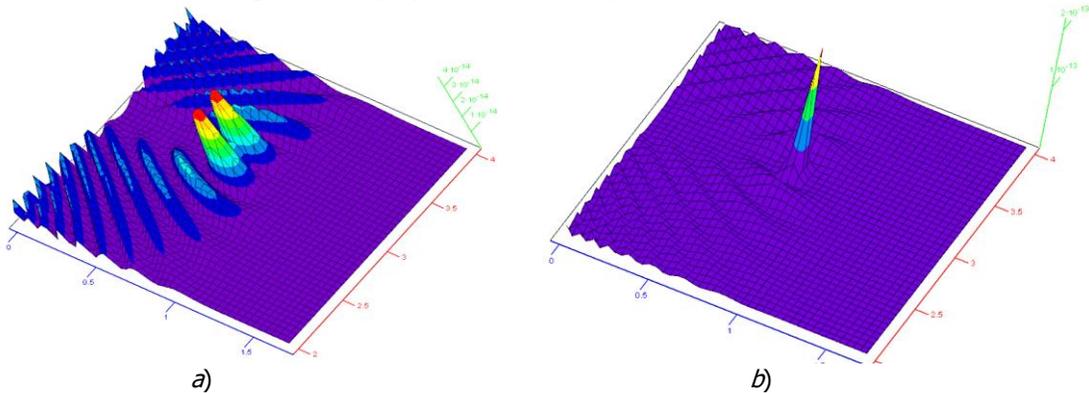

a)                                                                 b)

Fig. 19: The volumetric DESM on a multilayer non-isotropic crystal surface calculated by the formula (10) at $\vartheta = 45^0$, $\gamma = 0^0$, $l_1 = 10^{-11}$cm, $r_{cor5} = 4\times10^{-9}$cm, a) $n_1 = 48$ and b) $n_1 = 47$

If each layer of the crystal has the same anisotropy of the type (2.23a) in Batanov-Gaukhman (2020a), then for this case the following formula is obtained for calculating the DESM.

$$D(v,\omega/\vartheta,\gamma) = \frac{4}{\pi^2 l_2}\left(\frac{a^2 b^2}{(a^2+b^2)^2}\right)\left(\frac{\cos^2(\pi n_1) - \cos(\pi n_1)\cos\left(\sqrt{\frac{a^2+b^2}{d^2}}l_2/\eta\right)}{\left(\pi n_1/l_2\right)^2 - \left(\sqrt{\frac{a^2+b^2}{d^2}}l_2/\eta\right)^2} - \frac{\cos\left(\pi n_1 + \sqrt{\frac{a^2+b^2}{d^2}}l_2/\eta\right) - 1}{\left(\pi n_1/l_2 + \sqrt{\frac{a^2+b^2}{d^2}}/\eta\right)^2}\right) \times$$

$$\times \left|\frac{d(a'_\nu b'_\omega - a'_\omega b'_\nu) + c'_\nu(ba'_\omega - ab'_\omega)}{d^2\sqrt{a^2+b^2}}\right|. \quad (11)$$

where the scale parameter $\eta$ is given by expression (6).

The results of calculations using formula (11) are shown in Fig. 20.

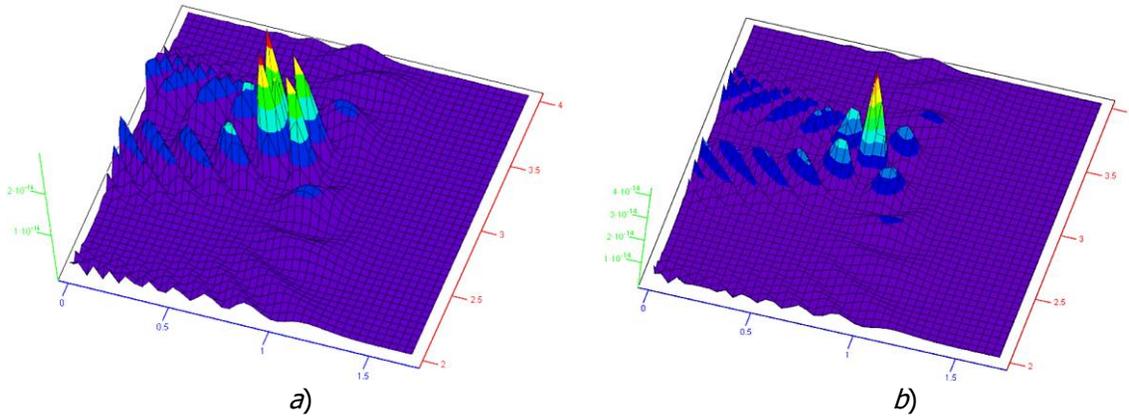

Fig. 20: The volumetric DESM on a multilayer non-isotropic crystal surface calculated by the formula (11) at $\vartheta = 45^0$, $\gamma = 0^0$, $l_1 = 10^{-11}$cm, $r_{cor5} = 4 \times 10^{-9}$cm, a) $n_1 = 42$ and b) $n_1 = 37$

## DISCUSSION

Extensive literature is devoted to the study of the diffraction of elementary particles (in particular, electrons and neutrons) and electromagnetic waves (in particular, gamma rays) by crystals of various substances.

Basically, the explanation and description of this phenomenon is based on the fact that the crystal is represented as a 3-dimensional periodic structure (3-dimensional atomic lattice), from which waves are reflected. In the case of laser or gamma rays, these are electromagnetic waves with the corresponding wavelength, and in the case of elementary particles, these are de Broglie waves with a wavelength (2). Within the framework of this method, the condition for the occurrence of interference of the reflected waves from atomic planes is determined by the Wolf-Bragg law (3); therefore, the phenomenon of wave scattering by a crystal is called Bragg diffraction.

Starting with the original works (Bragg, 1914) and Davisson & Germer (1928), this approach is used in most modern articles, for example Bragg (1914) and Davisson & Germer (1928). This approach is used in most modern articles, e.g., Bartell & Carroll (1965), Baudin *et al*. (1998), Bendersky & Gayle (2001), Bennett & Porteus (1961), Blazhevich *et al*. (2015), Bond (2016), Ching-Chuan (2002), Eisberg & Resnick (1986), Gehrenbeck (1978), Kuznetsov *et al*. (2015), McMahon (2013), Mikula *et al*. (2017), O'Donnell & Mendez (1987), Ogilvy & Merklinger (1991), Oleynikov *et al.* (2007), Pendry (1980), Prokes *et al.* (2019), Shi *et al.* (2013), Shintake (2021), Shorokhova & Kashin (2005), Suzuki & Suzuki (2013), Vainshtein (1964), Winkelmann & Vos (2011).

At the same time, there are works where the description of experiments on the diffraction of photons and electrons by a crystal is described in terms of the probability amplitudes and the space-time formulation of Feynman's quantum mechanics. This approach is proposed, for example, in the article Field (2013). There are also other methods of explaining Bragg diffraction, for example, in the article Logiurato *et al.* (2020), a geometric model is proposed to explain the pattern of wave diffraction on a two-dimensional grating.

The stochastic approach to studying the scattering of microparticles by statistically uneven layers of a crystal proposed in Batanov-Gaukhman (2020a) and in this article is completely different from all methods for solving this problem known to the author. According to the author, this approach makes it possible to explain the cause of the phenomenon under study on the basis of the principles of the classical theory of probability and statistical physics, i.e. without invoking de Broglie's hypothesis about the possible existence of the waves of matter. This method also makes it possible to calculate volumetric concentric scattering diagrams of microparticles on a crystal, which are in good agreement with experimental data. Whereas no other method of explaining Bragg diffraction known to the author allows calculating such diagrams.

**CONCLUSION**

Formulas (5), (10) and (11), obtained by the author in the article (Batanov-Gaukhman, 2020a, arXiv:2007.13527) and analyzed in this article, open up wide opportunities for studying the properties of multilayer crystal surfaces by analyzing the results of scattering of microparticles on them.

Selection of five parameters:

$$\vartheta^{H'}, \quad h_1^{V}, \quad n_1^{H}, \quad r_{cor}^{I}, \quad \gamma^{i}$$

which are associated with different properties of the atomic or molecular structure of the crystal, it is possible to achieve a similarity of the scattering diagram calculated by formula (5) {or (10) and (11)} with an electron diffraction pattern or an *X*-ray diffraction pattern, and thereby obtain information about their structure.

Calculations by formula (5) are in good agreement with the results of excrement on the diffraction of electrons and other microparticles on crystals, thin films and Davisson-Germer experiments (see Fig. 6 – 9). This allows us to propose a stochastic interpretation of these diffraction phenomena, described in the author's article (Batanov-Gaukhman, 2020a) and in this article. A stochastic explanation of these experiments can serve as an alternative to L. de Broglie's hypothesis about the existence of wave properties of matter.

In general, formulas (5), (10) and (11) with five parameters $\vartheta$, $h_1$, $n_1$, $r_{cor}$ and $\gamma$ generate an infinite set of two-dimensional surfaces (Bass-Fuchs worlds), in which individual forms can exist that reflect the outlines or essence processes in the surrounding reality.

At the same time, all these two-dimensional surfaces have a common property. Since formulas (5), (10) and (11) are the probability density functions, the double integral of all these functions over the angles v and $\omega$ from 0 to $2\pi$ is equal to unity

$$\int_0^{2\pi}\int_0^{2\pi} \rho(v,\omega/\vartheta,\gamma)|G_{v\omega}|\,dv\,d\omega = 1.$$

Formulas (5), (10) and (11) are suitable for describing the diffraction of not only elementary particles, atoms and photons, but also for the scattering of macroscopic elastic bodies (like a soccer ball) on large multilayer periodic structures.

Suppose, for example, a three-dimensional lattice is assembled from metal pipes with a diameter of 30 – 50 cm with an edge length of one cubic cell 3400 cm = 34 m, and metal balls with a diameter of 50 to 80 cm are placed in the nodes of this lattice (see Fig. 21).

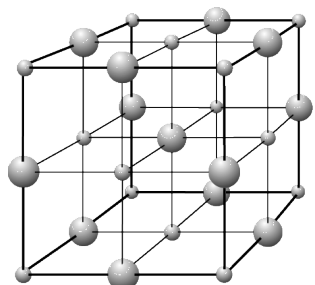

Fig. 21: A three-dimensional lattice is assembled from metal pipes and metal balls of various diameters

If a stream of the soccer balls with a diameter of 22.3 cm is directed to such a cubic lattice at a sliding angle $\vartheta = 45^0$, then their scattering is also described by formula (5). Indeed, if instead of $ƛ = 10^{-11}$cm, $r_{cor}= 6\times10^{-9}$cm and $n_1 = 66$ into the scale parameter $\eta$ (6) we substitute $ƛ = 50$ cm, $r_{cor}= 3400$ cm and и $n_1 = 18$, then the elastic scattering diagram of the soccer balls on such a cubic lattice, calculated by formula (5), will be approximately the same as shown in Fig. 6a.

If the case of diffraction of soccer balls is confirmed experimentally, then we can assert that formula (5) turned out to be universal with respect to various scales of the events under study, and the diffraction phenomena of the microcosm are indistinguishable from the phenomena of the macrocosm (under similar conditions).

Possible formulation of the inverse problem: - imitation of processes occurring in the microcosm, similar processes in the macrocosm. This will allow a more detailed understanding of the essence of microscopic phenomena.

In addition to solving practical problems, this article is aimed at bringing rational clarity to the mental problem associated with the discussion of the idea of the possible "existence" of de Broglie waves. The laws of geometric optics and probabilistic methods of statistical physics applied here made it possible, in the author's opinion, to explain the diffraction of elementary particles and atoms by crystals without invoking this hypothesis of Louis de Broglie. Moreover, in this work, it was suggested that the phenomenon of particle diffraction by solid periodic structures can manifest itself not only in the microcosm, but also in the macrocosm under similar conditions.

**ACKNOWLEDGEMENTS**

I thank my mentors, Dr. A.A. Kuznetsov and Dr. A.I. Kozlov for the formulation and discussion the problems outlined in this article. In performing the calculations, invaluable assistance was provided by Ph.D. S. V. Kostin. During the preparation of the manuscript, valuable remarks were made by D. Reed, Academician of the Russian Academy of Natural Sciences G.I. Shipov, Ph.D. V.A. Lukyanov and Ph.D. E. A. Gubarev, Ph.D. T.S. Levy.